\documentclass
[prl,twocolumn,tightenlines,showpacs,showkeys,10pt,final,floatfix]{revtex4}%
\usepackage{amsmath}
\usepackage{graphicx}
\usepackage{amsfonts}
\usepackage{amssymb}%
\begin{document}
\title{Spectroscopy of capacitively coupled Josephson-junction qubits}
\author{Philip R. Johnson}
\author{Frederick W. Strauch}
\author{Alex J. Dragt}
\author{Roberto C. Ramos}
\author{C. J. Lobb}
\author{J. R. Anderson}
\author{F. C. Wellstood}
\affiliation{Department of Physics, University of Maryland, College Park, Maryland 20742-4111}

\begin{abstract}
We show that two capacitively-coupled Josephson junctions, in the quantum
limit, form a simple coupled qubit system with effective coupling controlled
by the junction bias currents. We compute numerically the energy levels and
wave functions for the system, and show how these may be tuned to make optimal
qubits. The dependence of the energy levels on the parameters can be measured
spectroscopically, providing an important experimental test for the presence
of entangled multiqubit states in Josephson-junction based circuits.

\end{abstract}
\date{\today}
\maketitle

Ramos~\textit{et al.}~have proposed that electrically well-isolated Josephson
junctions can be used as qubits \cite{1}. Two recent experiments using
different isolation schemes have reported Rabi oscillations in single
junctions \cite{2}, demonstrating the existence of macroscopic quantum
coherence. While longer coherence times are desirable, these experiments show
that single Josephson junctions are strong candidates for solid-state qubits;
several Josephson-based types have been proposed \cite{3}.

One of the next major steps towards building a Josephson-junction based
quantum computer will be the observation of quantum properties of coupled
macroscopic qubits. A simple scheme for making coupled qubits, junctions
connected by capacitors, has recently been proposed by Blais \textit{et
al.~}\cite{4} and Ramos \textit{et al.}~\cite{5}. This scheme is illustrated
for the two-qubit case in Fig.~1(a).

\begin{figure}[tb]
\begin{center}
\includegraphics{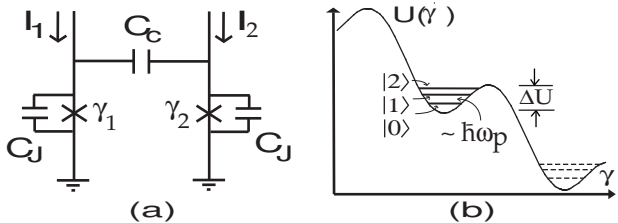}
\end{center}
\caption{(a) (left) Circuit diagram for two idealized capacitively coupled
Josephson junctions.~(b) (right) The tilted washboard potential for a single
current-biased Josephson junction with three metastable quantum states.}%
\end{figure}

In this paper, we focus on the immediately accessible fundamental
experiments--spectroscopic measurements of macroscopically entangled quantum
states--that are possible with this system. We calculate, using highly
accurate numerical methods, the energy levels and metastable wave functions
for the circuit of Fig.~1(a) in terms of the junction parameters, bias
currents, and coupling capacitance. Our numerical analysis demonstrates that
the system can be tuned to create appropriately spaced energy levels and
coupled states. The features that we discuss can both guide the experimental
effort of observing multiqubit quantum states and provide help in optimizing
the design of qubits and gates. But we emphatically stress that experimental
observation of these macroscopic entangled quantum states will be an important
achievement in its own right, and will provide strong support for the validity
of macroscopic quantum mechanics and the existence of macroscopic entanglement
\cite{6}. Spectroscopic observation of these states should be possible by
using standard single-junction experimental techniques \cite{5,7}.

The Hamiltonian for an ideal single current-biased Josephson junction, with
critical current $I_{c}$ and junction capacitance $C_{J},$ is
\begin{equation}
H\left(  \gamma,p\right)  =\left(  4E_{C}\hbar^{-2}\right)  p^{2}-E_{J}\left(
\cos\gamma+J\gamma\right)  ,
\end{equation}
where $\gamma$ is the gauge-invariant phase difference across the junction,
$J=I/I_{c},$ $I$ is the (tunable) bias current, $E_{J}=\left(  \Phi_{0}%
I_{c}/2\pi\right)  $ is the Josephson energy, $E_{C}=e^{2}/2C_{J}$ is the
charging energy, and $\Phi_{0}=h/2e$ is the flux quantum. The canonical
momentum is $p=\left(  \Phi_{0}/2\pi\right)  ^{2}C_{J}\dot{\gamma}=\hbar
Q/2e,$ where $Q$ is the charge on the junction$.$ The ratio $E_{J}/E_{C}$
determines whether the system is in a phase, charge, or intermediate regime.
The qubits explored in this paper have $E_{J}>>E_{C}$ and hence are in the
phase regime.

\begin{figure}[th]
\begin{center}
\includegraphics{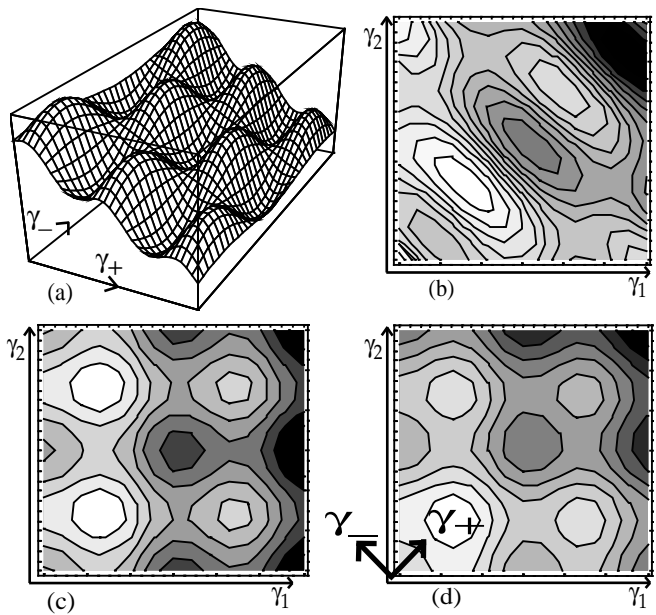}
\end{center}
\caption{(a) Potential $V^{\prime}$ with strong coupling ($\zeta=0.8$) and
$J_{1}=J_{2}$. The coupling induces a squeezing in the $\gamma_{+}$ direction,
relative to the $\gamma_{-}$ direction. (b) $V^{\prime}$ contours with
$\zeta=0.8$ and $J_{1}\neq J_{2}$, around the vicinity of one well. The
symmetry of $V^{\prime}$ shows that, despite detuned bias currents,
$\gamma_{+}$ and $\gamma_{-}$ are approximately the normal modes$.$ (c)
$V^{\prime}$ with small coupling ($\zeta=0.01$) and $J_{1}\neq J_{2}$ showing
(by symmetry) that $\gamma_{1}$ and $\gamma_{2}$ are approximate normal modes.
(d) $V^{\prime}$ with $\zeta=0.01$ and $J_{1}=J_{2}$ showing that $\gamma_{+}$
and $\gamma_{-}$ are normal modes.}%
\end{figure}

The equations of motion for a single current-biased junction are equivalent to
those for a particle in the tilted washboard potential shown in Fig.~1(b).
Classically, for $J<1$ there are stable minima about which the phase can
oscillate with the characteristic plasma frequency $\omega_{p}\left(
J\right)  =\sqrt{2\pi I_{c}/\Phi_{0}C_{J}}\left(  1-J^{2}\right)  ^{1/4}$
\cite{8}. Quantum mechanically, the system exhibits localized metastable
states in each well that can tunnel out into the running (finite-voltage)
state. The effective barrier height $\Delta U_{barrier}$ [see Fig.~1(b)] for a
single junction in units of $\hbar\omega_{p}\left(  J\right)  ,$ assuming
$J\lesssim1,$ is related to $N_{s}$ by%
\begin{equation}
N_{s}\simeq\frac{\Delta U_{barrier}}{\hbar\omega_{p}\left(  J\right)  }%
=\frac{2^{3/4}}{3}\left(  \frac{E_{J}}{E_{C}}\right)  ^{1/2}\left(
1-J\right)  ^{5/4}.
\end{equation}
Here, $N_{s}$ is the approximate number of metastable bound states for a
single isolated junction \cite{8}. Our analysis explores the relevant regime
for quantum computing where $N_{s}$ is small and the nonlinearity of the
Hamiltonian is important.

By adjusting the bias current it is possible to tune the barrier height to
obtain, for example, three metastable energy levels $E_{0}<E_{1}<E_{2}$ with
the two lowest states forming the basis $\left\vert 0\right\rangle ,\left\vert
1\right\rangle $ of a qubit. State $\left\vert 2\right\rangle $ has the
highest escape rate due to tunneling and can therefore act as an auxiliary
readout state, where readout is achieved by microwave pumping at a frequency
$\omega_{12}=\left(  E_{2}-E_{1}\right)  /\hbar.$ Detection of a voltage
across the junction implies that the system was previously in the state
$\left\vert 1\right\rangle $ and has entered the running state.

The Hamiltonian for the coupled two-junction circuit shown in Fig.~1(a) is
\begin{align}
H &  =\frac{4E_{C}}{\left(  1+\zeta\right)  \hbar^{2}}\left(  p_{1}^{2}%
+p_{2}^{2}+2\zeta p_{1}p_{2}\right)  \label{Two junction Hamiltonian}\\
&  -E_{J}\left(  \cos\gamma_{1}+\cos\gamma_{2}+J_{1}\gamma_{1}+J_{2}\gamma
_{2}\right)  ,\nonumber
\end{align}
where $\zeta=C_{C}/\left(  C_{C}+C_{J}\right)  $ is the dimensionless coupling
parameter, $C_{C}$ is the coupling capacitance, and $J_{1,2}$ are the
normalized bias currents of junctions 1 and 2, respectively. The canonical
momenta, $p_{1,2}=\left(  C_{C}+C_{J}\right)  \left(  \Phi_{0}/2\pi\right)
^{2}\left(  \dot{\gamma}_{1,2}-\zeta\dot{\gamma}_{2,1}\right)  ,$ are
proportional to the charges on each junction plus the charge on the coupling
capacitor plate adjacent to it \cite{4,5}.

By making a canonical change of variables, defined by
\begin{align}
\gamma_{\pm} &  =\left(  \gamma_{1}\pm\gamma_{2}\right)  /\sqrt{2\left(
1\pm\zeta\right)  },\label{gamma1,2 = gamma+,gamma-}\\
p_{\pm} &  =\sqrt{2\left(  1\pm\zeta\right)  }\left(  p_{1}\pm p_{2}\right)  ,
\end{align}
we find the transformed Hamiltonian%
\begin{equation}
H^{\prime}\left(  p_{+},p_{-},\gamma_{+},\gamma_{-}\right)  =\frac{4E_{C}%
}{\left(  1+\zeta\right)  \hbar^{2}}\left(  p_{+}^{2}+p_{-}^{2}\right)
+V^{\prime}(\gamma_{+},\gamma_{-}).
\end{equation}
Here, the momentum coupling term $2\zeta p_{1}p_{2}$ in the original
Hamiltonian is shifted to coupling in the new potential energy $V^{\prime
}=V^{\prime}(\gamma_{+},\gamma_{-}).$ Figures 2(a,b) illustrate how the
coupling induces a squeezing of $V^{\prime}$ along the $\gamma_{+}$ direction;
a strong coupling of $\zeta=0.8$ has been chosen to accentuate this behavior.

We gain further insight into the coupling of the junction states by looking at
the normal modes for small oscillations. For small coupling and detuned bias
currents $(J_{1}$ far from $J_{2}$) the normal modes are approximately
$\gamma_{1}$ and $\gamma_{2},$ i.e., the junctions are effectively decoupled.
This effect is shown in Fig.~2(c) by the approximate symmetry of $V^{\prime}$
with respect to reflections about the $\gamma_{1}$ and $\gamma_{2}$ axes. When
$J_{1}=$ $J_{2}$ the normal modes become $\gamma_{+}$ and $\gamma_{-},$ and we
therefore expect the coupled junction states to be entangled symmetric and
antisymmetric combinations of the single junction states. This can be seen in
Fig.~2(d), where $V^{\prime}$ is symmetric with respect to reflections about
the $\gamma_{+}$ and $\gamma_{-}$ axes. Figure 2(b) shows $V^{\prime}$ for
$\zeta=0.8$ and unequal bias currents; despite detuning the large $\zeta$
prevents the junctions from decoupling and $\gamma_{+}$ and $\gamma_{-}$
effectively remain as normal modes.

The challenging demands of quantum computing require an accurate and precise
quantitative description of the states going beyond simple perturbation
theory. To achieve this, we have computed the states and energy levels
numerically using a nonperturbative fast Fourier transform split-operator
method \cite{9} applied to the full nonlinear Hamiltonian in
Eq.~(\ref{Two junction Hamiltonian}). Our implementation computes the wave
functions on a lattice using a fourth-order integration of the imaginary-time
evolution operator $\exp(-\hat{H}\tau).$ While this method is relatively slow,
its results for a subset of system parameters confirm that the much faster
complex scaling method \cite{10} applied to the cubic approximation of the
full potential \cite{6} is accurate to at least 0.1\%. The faster complex
scaling method then allows us to compute energy levels for a wide range of
system parameters. A further important property of both these numerical
methods is that they are well suited to finding metastable states in
potentials that allow tunneling, particularly in more than one dimension,
where other methods fail. The computed quantum states have been further
verified by time evolving them on a lattice using real time split-operator
methods. This has confirmed dynamically that these states are truly
quasistationary and thus accurately determined; where applicable, agreement
has also been found with higher-order WKB analysis.

\begin{figure}[tb]
\begin{center}
\includegraphics{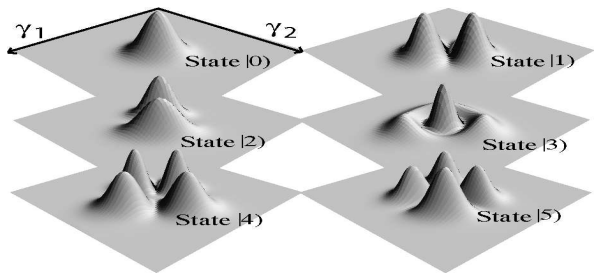}
\end{center}
\caption{The modulus squared of the probability amplitude of the first six
quasistationary wave functions for capacitively coupled current-biased
Josephson junctions with coupling of $\zeta=0.01$ and normalized bias currents
of $J_{1}=J_{2}=0.986\,93.$}%
\end{figure}

In Fig.~3 we show, for example, the numerically computed wave functions for
identical junctions with capacitances $C_{J}=4.3$ pF and critical currents
$I_{c}=13.3$ $\mu$A$.$ Junctions with these physical characteristics are
readily fabricated and of physical interest. Figure 3 shows the modulus
squared of the wavefunctions of the first six quasistationary states for the
coupling strength $\zeta=0.01,$ and with bias currents $J_{1}=J_{2}=0.986\,93$
such that isolated junctions would have approximately three quasistationary
states ($N_{s}\simeq3$). These large bias currents make the nonlinearities of
the potential pronounced, and the states deviate significantly from coupled
harmonic-oscillator states. The states $\left\vert n\right)  $ in Fig.~3 are
ordered by energy $E_{n}$; a rounded bracket has been used to distinguish the
coupled two-junction states $\left\vert n\right)  $ from single-junction
states $\left\vert n\right\rangle $. The second and third states expressed in
terms of single-junction direct product states are $\left\vert 1\right)
\cong\left(  \left\vert 01\right\rangle -\left\vert 10\right\rangle \right)
/\sqrt{2}$ and $\left\vert 2\right)  \cong\left(  \left\vert 01\right\rangle
+\left\vert 10\right\rangle \right)  /\sqrt{2}$, whereas the higher-energy
states are more complicated superpositions that depend upon the bias currents
and coupling. The ordering of the states in Fig.~3 may be understood by
looking at the potentials shown in Fig.~2(a,b); wave functions extended in the
$\gamma_{+}$ direction have higher energy because of the coupling induced
effective squeezing in the $\gamma_{+}$ direction, relative to the $\gamma
_{-}$ direction. Observe that because the $\left(  \gamma_{1},\gamma
_{2}\right)  $ configuration space variables are the collective degrees of
freedom of distinct junctions, the wave functions represent macroscopic
nonlocally correlated (and hence entangled) states.

\begin{figure}[tb]
\begin{center}
\includegraphics{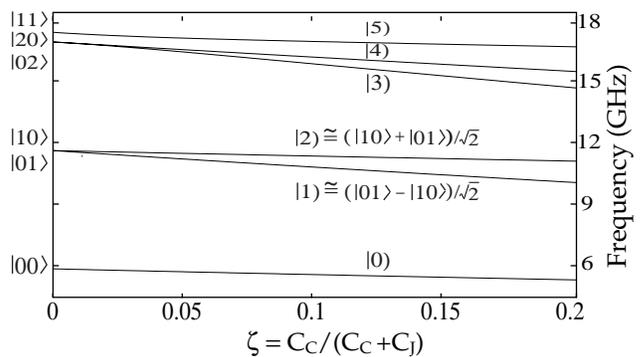}
\end{center}
\caption{Frequencies versus coupling strength for equal bias currents
$J_{1}=J_{2}=0.986\,93$, $I_{C}=13.3$ $\mu$A, and $C_{J}=4.3$ pF.}%
\end{figure}

For designing qubits out of coupled junctions we need to know how the energy
levels depend on coupling and bias current. Figure 4 shows the effects of
varying the coupling strength in the range $0<\zeta<0.2$ on the first six
energy levels with $J_{1}=J_{2}=0.986\,93.$ The plasma frequency of each
single junction when $\zeta=0$ is $\omega_{p}\left(  J_{1}=J_{2}\right)
/2\pi=6.2037$ GHz. The states are labeled at the left of Fig.~4 for zero
coupling, where the product representation $\left\vert nm\right\rangle
=\left\vert n\right\rangle \otimes\left\vert m\right\rangle $ is appropriate.
For zero coupling the nonlinearity of the potential has broken the degeneracy
between $\left\vert 5\right)  =\left\vert 11\right\rangle $ and the pair
($\left\vert 3\right)  =\left\vert 02\right\rangle ,\left\vert 4\right)
=\left\vert 20\right\rangle $).

\begin{figure}[tb]
\begin{center}
\includegraphics{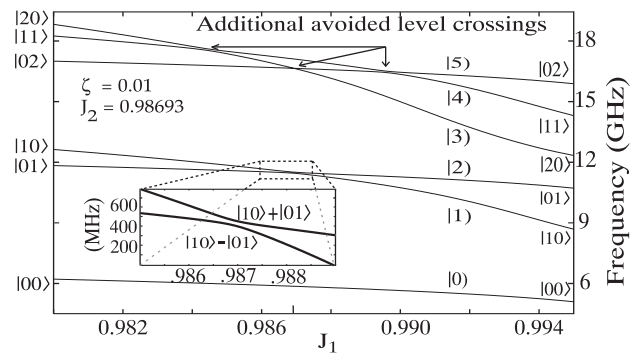}
\end{center}
\caption{Frequencies of the first six states versus bias current $J_{1}$ with
$J_{2}=0.986\,93$ fixed and a coupling strength of $\zeta=0.01$, $I_{C}=13.3$
$\mu$A, and $C_{J}=4.3$ pF.}%
\end{figure}

In real experiments, the coupling strength $\zeta$ will typically be fixed by
the circuit design, and can only be varied by making a completely new sample.
By contrast, the bias currents through each junction are easily varied, and
allow manipulation of the entangled states shown in Fig.~3. Figure 5 shows how
the energy levels change for $\zeta=0.01$ and $J_{2}=0.986\,93$ fixed, while
$J_{1}$ is varied around $J_{2}$. There are prominent avoided level crossings
indicated in the figure for both on-tune ($J_{1}=J_{2})$ and off-tune
($J_{1}\neq J_{2}$) bias currents. The predicted gap for the $\left\vert
1\right)  ,\left\vert 2\right)  $ on-tune splitting at $J_{1}=0.986\,93$ is
$57$ MHz for $I_{C}=13.3$ $\mu$A and $C_{J}=4.3$ pF. The predicted gap for the
first off-tune splitting between states $\left\vert 4\right)  $ and
$\left\vert 5\right)  $ at $J_{1}=0.984\,44$ is $80$ MHz, and for the second
off-tune splitting at $J_{1}=0.989\,62$ the gap is $72$ MHz. The predicted gap
for the on-tune $\left\vert 3\right)  ,\left\vert 4\right)  $ splitting of $4$
MHz is much smaller than the others because it is a second-order avoided
crossing in perturbation theory. Figure 6 shows the energy levels for the same
junction parameters but with $\zeta=0.05$ and $J_{2}=0.98$ ($N_{s}\simeq5$).

We have labeled the states in Figs.~5 and 6 at the far left and right--when
the currents are detuned and hence the states are effectively uncoupled--as
product states. This labeling is only strictly correct when $\zeta=0$. The
mixing that occurs between states when the bias currents are brought into tune
is indicated for the states $\left\vert 1\right)  $ and $\left\vert 2\right)
$ in Figs.~$4\ $and $5$. A swaplike gate operation can be constructed by
exploiting this mixing \cite{4}.

\begin{figure}[tb]
\begin{center}
\includegraphics{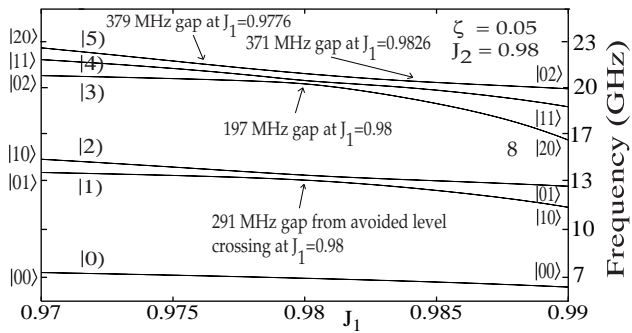}
\end{center}
\caption{Frequencies versus bias current $J_{1}$ with $J_{2}=0.98$ fixed and a
coupling strength of $\zeta=0.05,$ $I_{C}=13.3$ $\mu$A, and $C_{J}=4.3$ pF.}%
\end{figure}

Experimental data similar to Figs.~5 and 6 would be important first evidence
for the existence of macroscopic entangled states like those shown in Fig.~3.
A typical experiment to probe the energy levels in Figs.~4--6 would proceed by
preparing the system in the ground state $\left\vert 0\right)  $ by cooling
well below $T\simeq E_{01}/k\simeq300$ mK, where $E_{nm}=\left(  E_{n}%
-E_{m}\right)  ,$ and $E_{n}=E_{n}\left(  J_{i},\zeta\right)  $ are the
coupled-junction energies whose dependence on bias current and coupling we
have shown in Figs.~4--6. Varying the bias current $J_{1}$ (with $J_{2}$
fixed) while simultaneously injecting microwaves at a frequency $\bar{\omega}$
should lead to an enhancement in the tunneling from the zero-voltage state to
the finite-voltage running state of the system when $\bar{\omega}%
=E_{nm}\left(  J_{1}\right)  /\hbar.$ This enhancement produces a
corresponding peak in escape rate measurements \cite{5,7}. By varying
$\bar{\omega}$ and $J_{1}$ for the coupled junctions, we may map out the
energy levels for comparison with Figs.~5 and 6.

Experimentally, the expected energy gap between the avoided levels can be
resolved if both the quality factor $Q$ and coupling $\zeta$ of the system are
reasonably large. For example, with the typical junction parameters assumed
here, and with $\zeta=0.01$, the predicted gap for the $\left\vert 1\right)
$,$\left\vert 2\right)  $ splitting of $57$ MHz can easily be resolved with a
$Q$ of $200$ \cite{5}. Furthermore, one can reasonably track the bending of
the resonant escape peaks near the avoided crossings within the span of a
typical experimental current window of about 30 nA \cite{5}. These splittings
can be made even easier to detect by increasing the coupling capacitance (see
Fig.~6), though large coupling may inhibit efficient quantum gates.

In conclusion, we have presented predictions for fundamental experiments that
probe macroscopic entangled states by the relatively simple scheme of doing
spectroscopy on coupled junctions while varying external bias currents. The
energy levels of these entangled states should be readily observable using the
same experimental techniques that have allowed spectroscopy of single
junctions. The numerical methods we have used are powerful tools for mapping
out the metastable states of nonlinear, many-level coupled systems, and allow
us to explore a wide range of junction parameters and couplings. This kind of
detailed study will be necessary for the design of realistic coupled qubits.

\begin{acknowledgments}
We would like to thank Andrew J. Berkley, Huizhong Xu, and Mark A. Gubrud for
helpful discussions. This work was supported in part by the U.S. Department of
Defense and the State of Maryland, through the Center for Superconductivity Research.
\end{acknowledgments}


\begin{thebibliography}{99}                                                                                               %
\bibitem {1}R. Ramos \emph{et al}., IEEE Trans. Appl. Supercond. \textbf{11},
998 (2001).

\bibitem {2}Y. Yu \emph{et al.}, Science \textbf{296}, 889 (2002); J. M.
Martinis \emph{et al.}, Phys. Rev. Lett. \textbf{89}, 117901 (2002).

\bibitem {3}S. Han, R. Rouse, and J. E. Lukens, Phys. Rev. Lett. \textbf{76},
3404 (1996); M. F. Bocko, A. M. Herr, and M. J. Feldman, IEEE Trans. Appl.
Supercond. \textbf{7}, 3638 (1997); J. E. Mooij \emph{et al.}, Science
\textbf{285}, 1036 (1999); C. H. van der Wal \emph{et al.}, \emph{ibid.}
\textbf{290}, 773 (2000); J. R. Friedman \emph{et al.}, Nature (London)
\textbf{406}, 43 (2000); Y. Makhlin, G. Sch{\"{o}}n, and A. Shnirman, Rev.
Mod. Phys. \textbf{73}, 357 (2001); D. Vion \emph{et al.}, Science
\textbf{296}, 886 (2002).

\bibitem {4}A. Blais, A. Maassen van den Brink, and A. M. Zagoskin, condmat/0207112.

\bibitem {5}R. C. Ramos \emph{et al.}, IEEE Trans. Appl. Supercond. (to be
published) (2002).

\bibitem {6}A. J. Leggett, in \emph{Chance and Matter}, edited by J. Souletie,
J. Vannimenus, and R. Stora (Elsevier, Amsterdam, 1987), p. 395.

\bibitem {7}J. M. Martinis, M. H. Devoret, and J. Clarke, Phys. Rev. Lett.
\textbf{55}, 1543 (1985); Phys. Rev. B \textbf{35}, 4682 (1987).

\bibitem {8}T. A. Fulton and L. N. Dunkleberger, Phys. Rev. B \textbf{9}, 4760 (1974).

\bibitem {9}M. D. Feit, J. A. Fleck, Jr., and A. Steiger, J. Comput. Phys.
\textbf{47}, 412 (1982); J. E. Bayfield, \emph{Quantum Evolution} (Wiley, New
York, 1999).

\bibitem {10}R. Yaris \emph{et al.}, Phys. Rev. A \textbf{18}, 1816 (1978); E.
Caliceti, S. Graffi, and M. Maioli, Commun. Math. Phys. \textbf{75}, 51 (1980).
\end{thebibliography}
\end{document}